\documentclass[12pt,twoside]{article}
\usepackage{fleqn,espcrc1}

\usepackage{graphicx}
\usepackage[figuresright]{rotating}

\title{Type Ia supernovae: differences due to progenitors within delayed 
 detonation explosions} 

\author{I. Dom\'\i nguez\address{Universidad de Granada,  
        Granada, Spain}, 
        P. H\"oflich\address{University of Texas, 
                Austin, USA}
        and
        O. Straniero\address{Osservatorio Astronomico di Collurania, 
      Teramo, Italy}} 
       
\begin{document}

\maketitle

\begin{abstract}
 
At this moment, the use of SNIa for cosmology lies on the assumption that the 
SNe at high redshifts are equal to the local ones. However, some observations 
 indicate a correlation between light curve (LC)  properties and the morphological type 
  of the 
 host galaxy. This could indicate a dependence with the age (mass/composition) of the underlying 
 population. In this work we have chosen the delayed detonation explosion 
 model in CO Chandrasekhar mass WDs to explore the dependence of the SNIa 
 LC and nucleosynthesis with the initial mass and composition 
 of the WD progenitor. The progenitor influences the final SNIa via the 
 mass of the CO core formed and the C/O ratio within it (1D explosion 
 models). We have followed the evolution of stars with masses between 1.5 and 
 8 M$_\odot$ and metallicity, Z=0, 10$^{-5}$, 0.001 and 0.02, from the pre-main sequence to 
 the TP-AGB phase. The differences obtained in the final C/O ratio within 
 the explosive WD are smaller than 22$\%$.  
 This results in a difference at maximum of 0.03 mag and of 0.1 mag 
 when  
 the  brightness-decline relation is applied.   

\end{abstract}

\section{INTRODUCTION}

The increasing quality and quantity of Type Ia observations, indicate 
that they are not such an homogeneous class of events as it was previously 
thought. Moreover, several interesting correlations between observed  
 properties are obtained (see Leibundgut this volume).  These correlations 
   could give hints about the underlying explosion mechanisms 
 and progenitors. To understand the event and the reasons of the differences 
 is a pre-requisite to use SNIa for cosmology (at least, for us). The observations of SNIa 
 at high redshifts have lead to the conclusion that the expansion of the 
 Universe is accelerating \cite{Ri98,Pe99}. Together with the Cosmic Microwave 
 Background  observations, the mass density parameter and the cosmological 
constant have been estimated, $\Omega_{m}\sim$0.3 and $\Omega_{\lambda}\sim$0.7. 
 
There are some observational evidences that  indicate a relation 
 of SNIa with the morphological type of the host galaxy \cite{Br98,Wa97,Ha00}. 
 In ellipticals, as compare with
 spirals,   
 less events are found, they are more homogeneous (in maximum brightness 
 and decline rate of the light curve) and in general, they are dimmer.  
This could mean a correlation with the progenitor population and consequently  evolution in time.  

Different evolutionary paths could lead to an explosive degenerate WD  and 
 several explosion mechanisms could blow up this WD (see Woosley this volume).
 Probably each of them contribute {\em up to some degree} to the SNIa population.  In this work 
 we have decided to fix both, the scenario and explosion mechanism and just change 
 the initial mass and composition of the WD progenitor.    
   
We have chosen the delayed detonation (1D) explosion of a Chandrasekhar mass 
 WD because this scenario accounts for 
most of the observational constraints. The key parameter is the transition 
 density, $\rho_{tr}$, the density at which the deflagration turns into 
a detonation. Varying $\rho_{tr}$ the   
optical and IR light curves and
spectra evolution of normal and subluminous SNIa are reproduced. Subluminous 
 SNIa are found to be redder,  as observed. Moreover, the  
brightness-decline relation (LCs of brighter events decline slower) used  
 in all the cosmological applications, is 
 obtained \cite{Ho95,Ho96}.
 Other correlations, like the  
CaII H+K  $\&$ Si minimum velocities and mean Ni velocity with brightness 
 are as well reproduced.

\section{MODELS}

We have studied the evolution of intermediate mass stars, with masses 
 in the range, 1.5 $\leq$ M/ M$_\odot$ $\leq$ 8, and initial composition,  
Z=0, 10$^{-5}$, 0.001 (Y=0.23) and Z=0.02 (Y=0.285). The evolution is 
 followed from the pre-main sequence to the thermal pulse AGB phase, 
including several pulses. The mass and chemical structure of the obtained 
 degenerate 
CO core depends on the initial mass and composition. It is assumed  that the
   {\em future} CO white dwarf has the mass and chemical structure of this  
 CO core. Accretion   
of H on the WD is performed at high rates up to the central C ignition at $\rho_c$=2.0 10$^9$ g/cm$^3$. At that time, the mass of the CO WD is close to 1.37 M$_\odot$. Note that the final amount of C and O in the accreted matter is 
 nearly equal (C/O$\approx$1). 

For the set with Z=0.001, delayed detonation explosions, detailed 
 post-processing and light curves are computed. 
The description of the velocity of the deflagration front is  based on 3D 
simulations \cite{Do00}. Explosions with several transition 
densities are performed,  
$\rho_{tr}$: 1.5,  1.8,  2.0,  2.3,  2.5  and 2.7 10$^7$ g/cm$^3$.

\subsection{Numerical methods}

The 1D hydrostatic evolutionary code FRANEC is used for the stellar evolution \cite{Chi98}. 
An extended nuclear network is employed for both, the H burning (269 reactions) and 
 the He burning (147 reactions). A time dependent mixing is adopted and the 
 physical and chemical evolutions are coupled. 

The explosions and light curves are computed with a 1D radiation-hydrodynamic 
 code including nuclear networks \cite{Ho96}. The hydrodynamics 
 equations are solved explicitly by the piecewise parabolic method \cite{Co84},  the frequency averaged radiation transport equations are solved implicitly via moment equations and expansion opacities and a detailed equation of state are 
 included.  
  The nuclear burning (post-processing) includes 218 to 606 isotopes 
 \cite{Thi96}. The same code is used for the subsequent expansion and bolometric and monochromatic light curves calculations. In this phase nuclear burning 
 is neglected and  
 gamma ray transport is taken into account using a Monte Carlo scheme. 
 The scattering, photon redistribution and thermalization terms used in the light curve opacity calculation are calibrated with NLTE calculations \cite{Ho95}.

\section{RESULTS}

\begin{figure}[tbp]
\centerline{
\includegraphics[angle=270,scale=0.4]{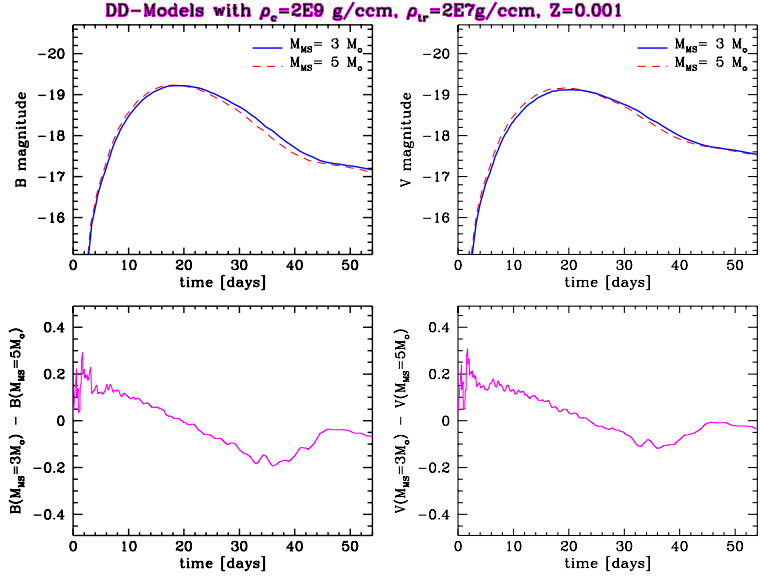}}
\caption[]{\small Comparison of the B and V light curves of models 
 with different progenitor initial masses, 3 M$_\odot$ and 5 M$_\odot$, and otherwise identical 
parameters}
\end{figure}
We do not expect  variations beyond a few tenths of a magnitude 
by  changing the initial mass and composition of the progenitor. Previous
studies were based on parameterized C/O ratios \cite{Ho98}
 or a  progenitor with  $7M_\odot$ \cite{Ho00}  indicating a change of
$0.3^m$ for a 40 \% change in the C/O or $0.1^m$ for a change in the metallicity from 0.02 to 0.001, respectively.
In our calculations, the maximum difference  
 obtained in C/O is $\leq$ 22$\%$. Note that the amount of accreted matter (bigger for the smaller cores) acts 
 as a uniform factor. This means that in the case of {\em pure} mergers (two cores) 
 the final difference would be slightly bigger.

  C/O is more sensitive to the initial mass than to the 
  initial composition. For Z=0.001,  we have followed the explosion and we have computed the LCs.  
  Varying the   
 initial mass from 3 to 5 M$_\odot$,  changes C/O from 0.78 to 0.73 M$_\odot$, 
    Ni mass from 
 0.51 to 0.48 M$_\odot$ (6$\%$ !) and kinetic energy from 1.18 to 1.17 10$^{51}$ erg (for $\rho_{tr}$= 2.0 10$^7$ g/cm$^3$). 
 In Figure 1, we show the  B and V light curves for  
 the 3M$_\odot$ and 5M$_\odot$ models.

 In summary, M$_{MAX}$ is 
 found to change from 0 to 
 0.03 mag when the initial mass and composition of the WD progenitor 
 vary.  The application of the 
brightness-decline relation results in  $\Delta M_{MAX}\leq$ 0.1 mag. 

Greater differences in the LCs are 
likely produce by the burning conditions, as $\rho_{tr}$ \cite{Ho95}. For example, 
 for the 3M$_\odot$, Z=0.001 models,  
  a change in  $\rho_{tr}$ from 1.5 to 2.7 10$^7$ g/cm$^3$ leads to 
 a change in the Ni mass from 0.17 to 0.72 M$_\odot$. We like to stress that 
 we treat progenitors and burning conditions completely uncoupled and 
 this is probably not the case. In 3D calculations the composition gradients 
 would influence the velocity of the front. Other effects like mixing and 
 rising blobs could be relevant. On the other hand,  
  different scenarios,     
 rotation and crystallization could strongly modified pre-explosive conditions. 
    
 Other important constraints obtained from the observations are the velocity 
range in which the different elements are found and the isotopic abundances 
 as compare with the solar abundances. 
 For the same transition density, differences in the global kinetic energy 
 as function of mass and composition of the progenitor are small and the 
 velocity range for the different elements does not vary in more than 
 few hundreds km/s. $^{68}$Zn, $^{48}$V and $^{4}$He change by more than 
 a factor of 5 when initial mass is changed. While a change greater than 
 a factor 2 is found for $^{14}$N, $^{17}$O, $^{18}$O, $^{19}$F, $^{20}$Ne,  $^{21}$Ne, $^{23}$Na, $^{25}$Mg, $^{26}$Mg, $^{42}$Ca, $^{43}$Ca, $^{46}$Ti, $^{60}$Ni, $^{61}$Ni, $^{64}$Zn, $^{66}$Zn, $^{67}$Zn, $^{63}$Cu and $^{65}$Cu.  

 As a final remark, we like to stress that    
  the final CO core mass and C/O ratio depend on the 
  treatment of turbulent convection  and on the $^{12}C(\alpha,\gamma)^{16}O$ reaction rate \cite{Do99} (see Imbriani et al., this volume). Changing this reaction rate from the high  
  to the low rate given 
 by \cite{Bu96}, remarkable differences in C/O and consequently  
 in  Ni mass and kinetic 
 energy are obtained. For the 3 M$_\odot$, Z=0.001 C/O changes
 from 0.74 to 1.53;  
Ni mass from 0.50 to 0.63 M$_\odot$  and the kinetic energy from 1.15 to 1.37 
 10$^{51}$ erg ($\rho_{tr}$=2.0 10$^7$ g/cm$^3$).   


\end{document}